\documentclass[twocolumn,showpacs,preprintnumbers,amsmath,amssymb]{revtex4}
\usepackage{epsfig}
\usepackage{amsmath}

\begin{document}

\title{Diffusion of Time-Varying Signals in Complex Networks: 
A Structure-Dynamics Investigation Focusing the Distance to the Source
of Activation}

\author{Luciano da Fontoura Costa}
\affiliation{Institute of Physics at S\~ao Carlos, University of
S\~ao Paulo, PO Box 369, S\~ao Carlos, S\~ao Paulo, 13560-970 Brazil}

\date{3rd Nov 2008}

\begin{abstract}
The way in which different types of dynamics unfold in complex
networks is intrinsically related to the propagation of activation
along nodes, which is strongly affected by the network connectivity.
In this work we investigate to which extent a time-varying signal
emanating from a specific node is modified as it diffuses, at the
equilibrium regime, along uniformly random (Erd\H{o}s-R\'enyi) and
scale-free (Barab\'asi-Albert) networks.  The degree of preservation
is quantified in terms of the Pearson cross-correlation between the
original signal and the derivative of the signals appearing at each
node along time. Several interesting results are reported.  First, the
fact that quite distinct signals are typically obtained at different
nodes in the considered networks implies mean-field approaches to be
completely inadequate.  It has also been found that the peak and lag
of the similarity time-signatures obtained for each specific node are
strongly related to the respective distance between that node and the
source node.  Such a relationship tends to decrease with the average
degree of the networks.  Also, in the case of the lag, a less intense
relationship is verified for scale-free networks.  No relationship was
found between the dispersion of the similarity signature and the
distance to the source.
\end{abstract}

\maketitle

\vspace{0.5cm}
\emph{`You cannot conceive the many without the one.' (Plato)}

\section{Introduction} 

One of the most important current issues in complex systems research
regards the identification of relationships between the connectivity
structure of a given network and the properties of different types of
dynamics (e.g. synchronization, Ising and integrate-and-fire)
unfolding on that network, giving rise to the so-called
structure-dynamics paradigm.  The present work focuses on the
investigation of the alterations of activation signals as they
diffuse, at the equilibrium regime, along networks.  More
specifically, a given node is selected as the source of the
activation, which is subsequently diffused among the neighboring
nodes, allowing the signals arising at each node to be compared to the
original activation.  The choice of such a linear dynamics monitored
with respect to pairs of nodes (i.e. the source and the node where the
signal is observed) is justified for the following reasons: (i) as
diffusion plays an important role in many linear systems,
investigations of signal diffusion can provide valuable initial
insights about a large number of natural and artificial systems; (ii)
because diffusion dynamics is involved in many non-linear dynamics
(e.g. reaction-diffusion), such studies can yield preliminary insights
about those more sophisticate dynamics; and (iii) the consideration of
the activations at individual nodes provides a much more comprehensive
characterization of the structure-dynamics relationship than typical
mean-field approximations, which are only valid when the activations
are very similar amongst nodes.

The conservation along time of the activation injected into the
network implies the signals at the nodes to progressively increase.
In order to avoid this trend, we consider the first time derivative of
the signals at each node, which guarantees null signal average.  Thus,
the transformation undergone along time as the activation is
transported from the source to a given node $i$ can be quantified in
terms of the Pearson correlation coefficient between the derivative of
the signal arising at that node and the original activation signal.

The Pearson correlation, instead of the traditional correlation or
covariance, is adopted here in order to normalize the magnitude of the
obtained similarity signals, which becomes comprised between -1 and
1. This also removes the effect of the respective magnitudes of the
two signals being compared.  The similarity between the observed
signal and the original signal injected at the source allows several
interpretations.  Its maximum absolute value (peak) reflects how
intensely the original signal appears within the observed node.  At
the same time, the time where such a peak appears indicates the
relative delay of the observed signal with respect to the source.
Observe that a high value of peak magnitude does not necessarily means
that the original signal is accurately reproduced at the observed
node, as large dispersions around the peak suggests cluttered versions
of the original signal.  Therefore, we also quantify such a dispersion
in terms of the entropy of the Pearson correlation.

In order to provide a first glimpse about how the network structure
influences the diffusion of time signals, we briefly discuss a simple
case example.  Figure~\ref{fig:ex1}(a-b) shows two networks receiving
activation from the respective source node (node number 1 in both
cases).  The activation consists of just a single Kronecker delta of
intensity 2 at time step $t=1$.  For simplicity's sake, we assume that
the signal diffusion is related to self-avoiding random walks, a
non-linear dynamics closely related to traditional random walks.  Let
us focus on the activation arising at node 3. In
Figure~\ref{fig:ex1}(a), the signal arising at that node will consist
simply of the original signal delayed by 1 time-step.  However,
because of the additional path connecting nodes 1 and 3 in the network
in Figure~\ref{fig:ex1}(b), the signal resulting at node 3 will be a
linear combination (2 terms) of delayed versions of the original
signal.

It is clear from the example above that the effect of the network
structure on the propagation of the activation signals along the
network is strongly related to its topology, in this case to the
number of paths of different lengths existing between the source and
each other individual node.  Therefore, every network characterized by
varying path statistics cannot be properly described or analyzed in
terms of mean-field approximations of the signals taken along the
whole network. Though implying in different dynamics, where portions
of the signal are send backwards at each step, and at the equilibrium
regime, the linear diffusion (traditional random walks) studied in
this work can be expected to be affected in similar ways by the
network structure.

\begin{figure}
  \vspace{0.3cm}
  \centerline{\includegraphics[width=1\linewidth]{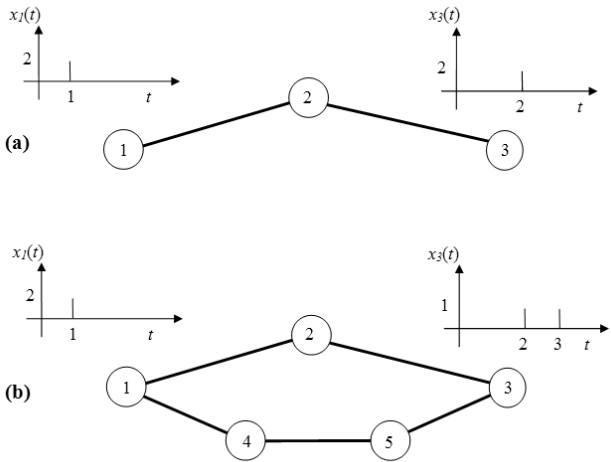}} \caption{A
  signal consisting of a Kronecker delta is diffused through self-avoiding
  random walks from the source node (i.e. node 1) to other nodes in
  two different networks (a and b), while the activation resulting at
  node 3 is observed along the initial time steps (transient
  dynamics).  In the case shown in (a), the effect of the network is
  only to delay the appearance of the signal at node 3.  In (b), the
  existence of two alternative paths with different lengths 
  extending from node 1 to 3 imply
  in widening the signal observed at the latter node. This figures
  illustrates the fact that the number of paths of different lengths
  between each node and the source plays a decisive role in modifying
  the activation signal as it propagates through the network.}  
  \label{fig:ex1}
\end{figure}

The current work investigates the propagation of time-signals through
Erd\H{o}s-R\'enyi (ER) and Barab\'asi-Albert (BA) complex networks, at
the equilibrium regime, as implied by the linear dynamical process of
random walks.  The signals observed at each node are differentiated
along time and compared to the original activation by using the
Pearson correlation between those signals.  Three related measurements
are taken into account: the magnitude of the correlation peak, its
position along time (lag), as well as the correlation dispersion
(quantified in terms of entropy).  These experiments are performed
with respect to ER and BA networks with fixed size (200 nodes) and two
average degrees, namely 2 and 4. The consideration of these two
theoretical complex network models allows us to investigate how
different types of connectivity affect the diffusion of the
time-varying signals. A number of interesting results are obtained and
discussed, including the identification of well-defined relationships
between the peak magnitude, lag and dispersion with the respective
distance to the source node.

This work starts by presenting the basic concepts related to complex
networks representation and characterization, as well as simulation of
the diffusion dynamics, synthesis of the activation signals, and
Pearson correlation between two signals.  A simple example is then
presented in order to clarify these concepts and obtain some initial
insights about how the structure of complex networks influences the
propagation of time-varying signals.  The results are subsequently
presented and discussed, and suggestions for future related research
are proposed.

\section{Basic Concepts}

A complex network is a graph exhibiting a particularly complex
structure.  The current work focuses unweighted, undirected networks.
A network of this type can be fully represented in terms of its
\emph{adjacency matrix} $K$, such that each undirected edge between nodes $i$
and $j$, with $i,j \in \{1, 2, \ldots, N \}$, implies
$K(i,j)=K(j,i)=1$, with $K(i,j)=K(j,i)=0$ being enforced otherwise.
Observe that the adjacency matrix $K$ has dimension $N \times N$.  The
\emph{degree} of a node $i$, hence $k(i)$, corresponds to the number
of edges attached to it.  Two nodes are adjacent if they share
different extremities of an edge.  Two edges are adjacent if they
share a single node.  A walk is a sequence of adjacent edges or nodes.
A \emph{path} is a walk where nodes are edges cannot be repeated. The
length of a path is henceforth understood to mean the number of its
constituent edges. The shortest path between two nodes corresponds to
the path comprised between those nodes which contains the smallest
number of edges.  We consider two theoretical models of networks,
namely the uniformly random Erd\H{o}s-R\'enyi and the scale-free
Barab\'asi-Albert structures (e.g.\cite{Albert_Barab:2002,
Costa_surv:2007, Newman:2003}), which are generated as described
in~\cite{Costa_surv:2007}.

The diffusion of activations in a complex network can be effectively
obtained in terms of its respective \emph{transition matrix} $S$,
which can be obtained from the adjacency matrix as follows

\begin{equation}
  S(i,j) = \frac{K(i,j)}{\sum_{j=1}^{N} K(i,j)},
\end{equation}

so that the sum along any of the columns of $S$ yields 1 as a
result. The activation at each of the $N$ nodes at a given time $t$
can be represented in terms of the state vector

\begin{equation}
  \vec{v}(t) = (v_1(t), v_2(t), \ldots, v_N(t))^T.
\end{equation}

so that $v_i(t)$ is the state at node $i$.  Thus, given the network
state at a specific time $t$, its state at any subsequent time
$t+\Delta t$, where $\Delta t = 0, 1, \ldots$, the linear
dynamics of diffusion can be calculated as

\begin{equation}
  \vec{v}(t+\Delta t) = S^{\Delta t} \vec{v}(t) + \vec{s}(t),
\end{equation}

where each component $i$ of $\vec{s}(t)$, i.e. $s_i(t)$, represents a
forcing signal arriving at node $i$.  In this work, just one of the
nodes in the network, henceforth called the \emph{source}, receives a
non-zero forcing signal along time.

All activation signals propagated from the source node in this work
are uniformly random, Bernoulli realizations with probability
$\alpha$.  Thus, a forcing signal $s(t)$ of duration $L$ can be
generated by starting with a vector of $L$ zeros and then changing
each of the zeros to one with fixed probability $\alpha$.  During the
investigations reported in Section~\ref{sec:exp} of this work, $8$
periods of the random input signal are fed into the system, and the
Pearson correlation is calculated only for the penultimate period,
when the dynamics is already at the equilibrium regime.

In order to remove the progressive increase of the activations along
time which will be otherwise observed at each node as a consequence of
the conservation of the activation implied by traditional random
walks, we take into account in our analysis the first time derivative
of those signals.  Because we are using discrete time signals with
unit time step, their time derivative can be obtained by subtracting
the value of the signal at the current time from the value at the
previous time.

The similarity between any two time-varying signals of the same length
can be quantified in several ways.  In this work, we resort to the
Pearson correlation coefficient between two signals.  Let $a(t)$ and
$b(t)$, with $t = 1, 2, \ldots, L$, be two time-varying signals.
First we standardize each of these signals.  In the case of signal
$a(t)$, its standardization can be calculated by making, for each $t$,

\begin{equation}
  A(t) = \frac{a(t) - \left< a \right>}{\sigma_{a}},
\end{equation}

where $\left< a \right>$ is the time-average of $a(t)$ and $\sigma_{a}$
is the standard deviation of the values of $a(t)$.  The Pearson
correlation coefficient between the two signals can now be calculated
as

\begin{equation}
  P_{a,b}(t) = \frac{1}{L-1} \sum_{i=1}^{L-t+1} A(i) B(i+t-1)
\end{equation}

The Pearson correlation coefficient resulting signal, i.e. $P(t)$, has
values which are necessarily between -1 and 1.  Values close to 1
indicate that one of the signals is almost completely contained into
the other signal at that lag $t$.  Contrariwise, values close to 0
mean that the two signals are quite different for that lag. Values
near to $-1$ indicate that the a similar version of the original
signal appears with negative sign.  Observe also that the preliminary
standardizations of both signals ensure that the correlation becomes
invariant to multiplications of any of the signals by a constant, as
well as respective additions with a constant.  The Pearson correlation
coefficients between two time-varing signals of length $L$ is also a
time-varying signal of length $L$, which in this work is called
similarity signature.

In order to characterize the similarity between the original signal
and the derivatives of the signals arising at each of the nodes in a
given network, we resort to the following three features extracted
from the Pearson correlation signal: (i) the maxium absolute value
('peak') of the Pearson correlation, which is proportional to the
maximum overlap between the two signals being compared; (ii) the
distance from the time where the peak occurs and the origin of the
respective period, which corresponds to the delay in which the maximum
similarity is obtained; and (iii) the entropy of a whole Pearson
correlation period, which quantifies how dispersed the identified
signal is.  The latter measurement is obtained by understanding the
Pearson correlation period as a probability density (it is therefore
normalized so that its overall sum is 1) and using the classical
definition of entropy.  Thus, low entropy indicates that the
derivative of the signal at the respective node contains a relatively
uncluttered version of the original signal within itself.

\subsection{A Simple Example of Signal Diffusion}

\begin{figure}
  \vspace{0.3cm}
  \centerline{\includegraphics[width=0.9\linewidth]{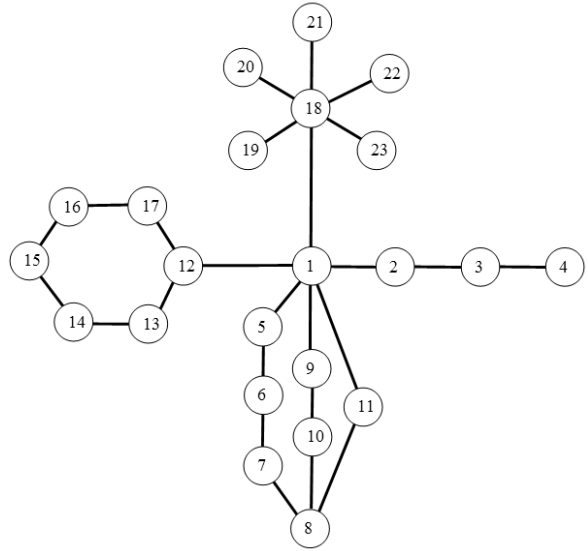}}
  \caption{A simple network used to illustrate the diffusion of
           an activation signal emanating from node 1.}
  \label{fig:net2}
\end{figure}

Let us illustrate the above concepts with respect to the simple
network shown in Figure~\ref{fig:net2}.  This network is composed of
four main branches deriving from node 1, each one representing a
different type of connectivity: (a) a chain of nodes extending from
node 1 to node 4; (b) three alternative paths with different lengths
between node 1 and 8; (c) a cycle with 6 edges including node 15; and
(d) node 1 connected to a hub (node 18).

Figure~\ref{fig:ex1} shows three periods (each with $L=200$ time
steps) of a random signal with $\alpha = 0.05$ (a), injected at node 1
of the network in Figure~\ref{fig:net2}, as well as the time
derivatives of the signals appearing at nodes 4, 8, 15 and 18,
followed by the respective similarity signatures with the injected
signal, as quantified by the Pearson correlation.  It can be observed
from all the signal derivatives that they are composed by a linear
combination of a basic signal packet corresponding to the derivative
of the system response to a single impulse.  In other words, the
signal derivative at a given node can be obtained by convolving the
original signal with the derivative of the system response obtained
for that node.  Observe also that the latter signal involves
alternance of positive and negative values, so that their linear
combination can lead to increase or decrease of the resulting signal,
depending on the interval between successive input pulses.  As a
consequence of the small size of the network in Figure~\ref{fig:net2},
the transient dynamics can hardly be discerned and is limited to the
first period.  Because of the finite duration of the signals, the last
period in the Pearson coefficients shows a decay as it approaches the
final time steps.  Observe also that the Pearson correlation
coefficients (indicated as 'Whole Pearson' in the figure) for each
node always exhibit a peak near the beginning of each period.  The
magnitude and lag, measured with respect to the beginning of each
period, of the peaks resulted similar for nodes 4, 15 and 18 (just
below 0.04) and more than doubled (0.085) for node 8.  The lags
obtained for nodes 4, 8, 15 and 18 were 2, 1, 3, and 2 time steps,
respectively.  Except for node 18, these lags tended to reflect the
distance from the respective nodes to the source of activation.
Another important aspect of the similarities, quantified in terms of
the Pearson correlation coefficients, concerns the dispersion of
activations around the peak, a feature which is strongly affected by
the dispersion of the system impulse response.  Largest dispersions
can be observed for nodes 15 and 18, followed by nodes 4 and 8.  The
smaller dispersion and higher magnitude of the peak obtained for the
latter case are to a substantial extent a consequence of the path of
length 2 edges interconnecting node 8 to the source, as well as the
fact that more activation is driven from the source to that node
through the three paths going from the source to node 8.

\begin{figure*}
  \vspace{0.3cm}
  \centerline{\includegraphics[width=0.9\linewidth]{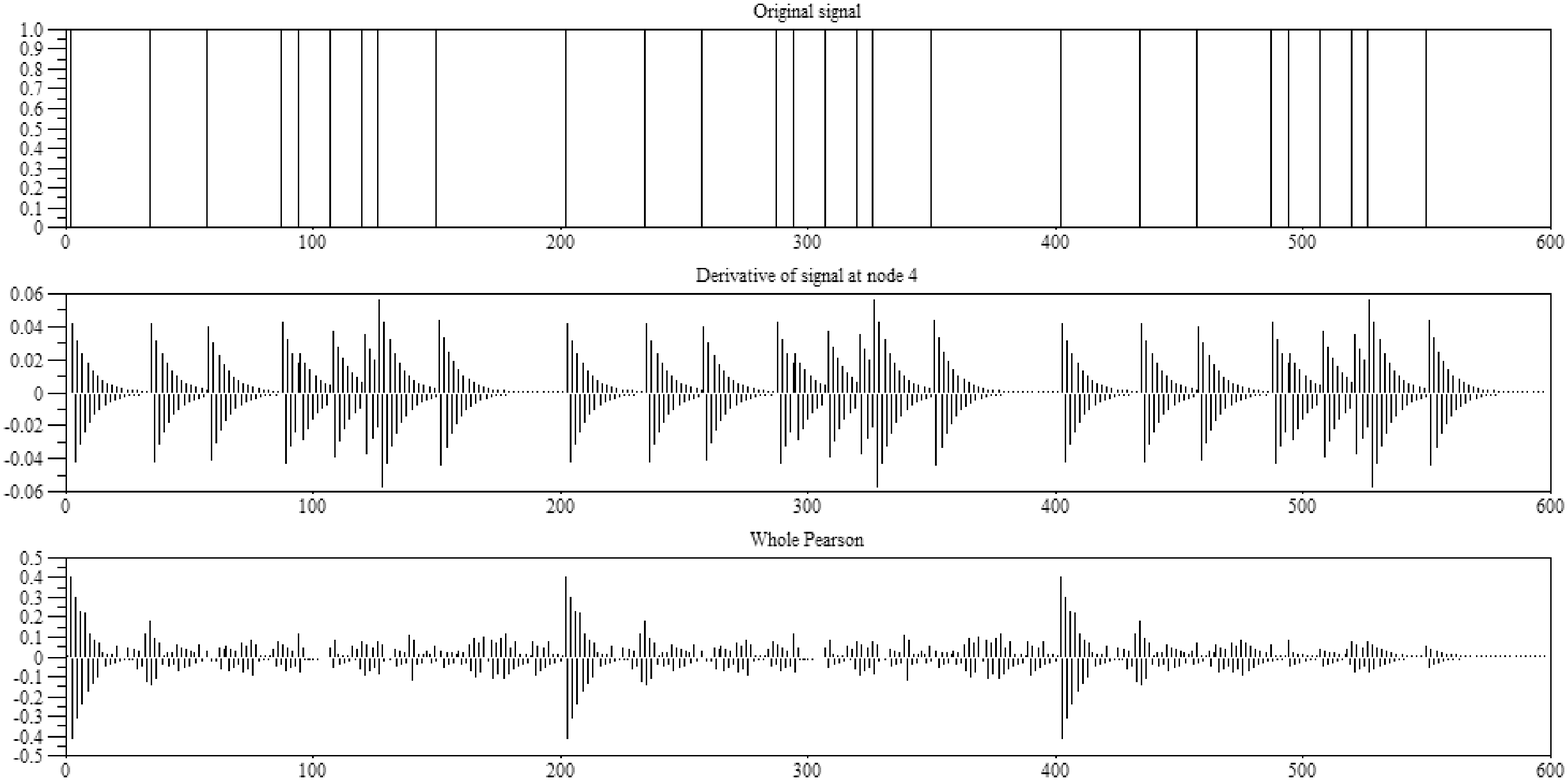}}
  \centerline{\includegraphics[width=0.9\linewidth]{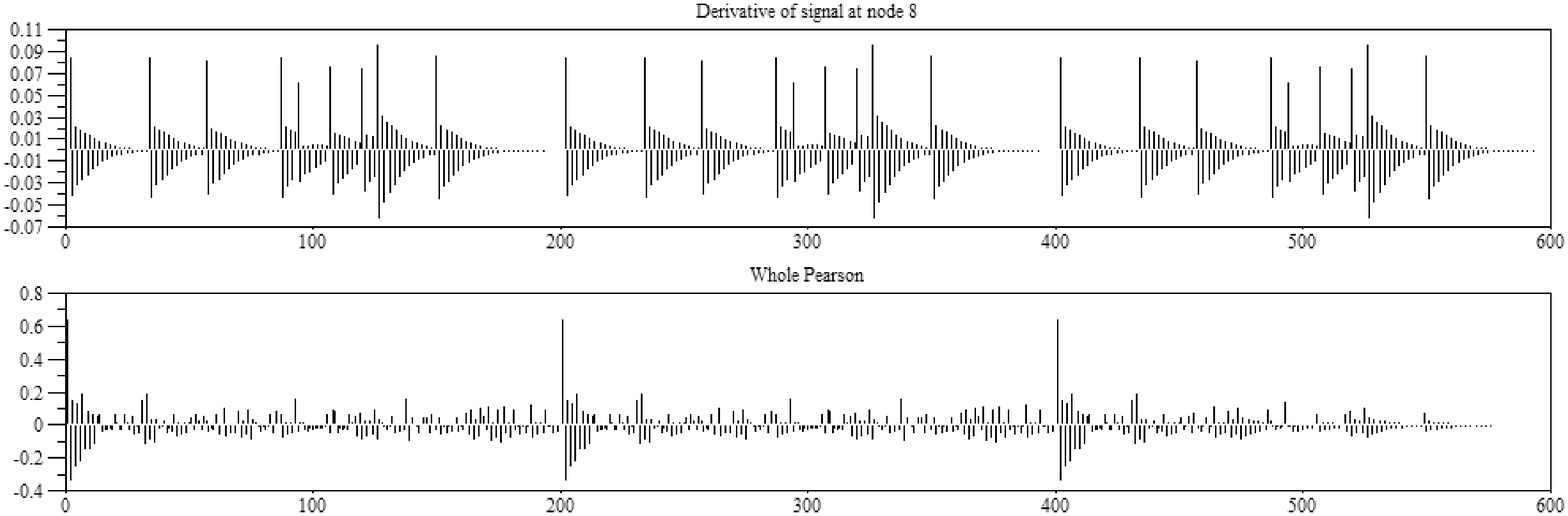}}
  \centerline{\includegraphics[width=0.9\linewidth]{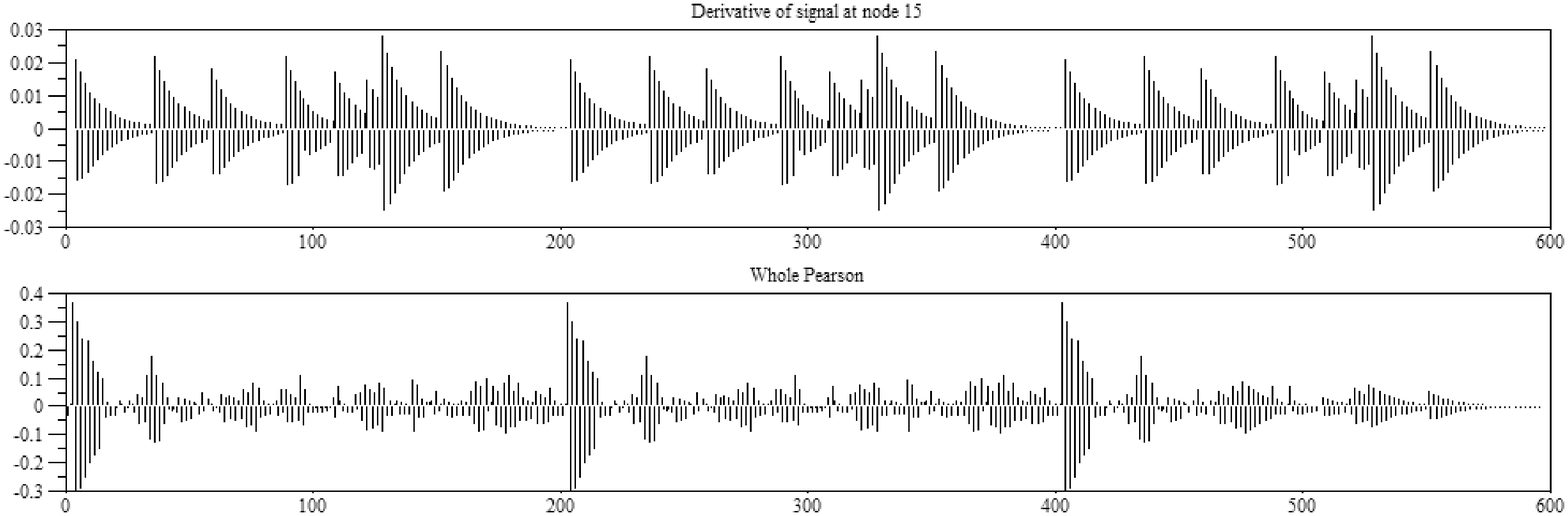}}
  \centerline{\includegraphics[width=0.9\linewidth]{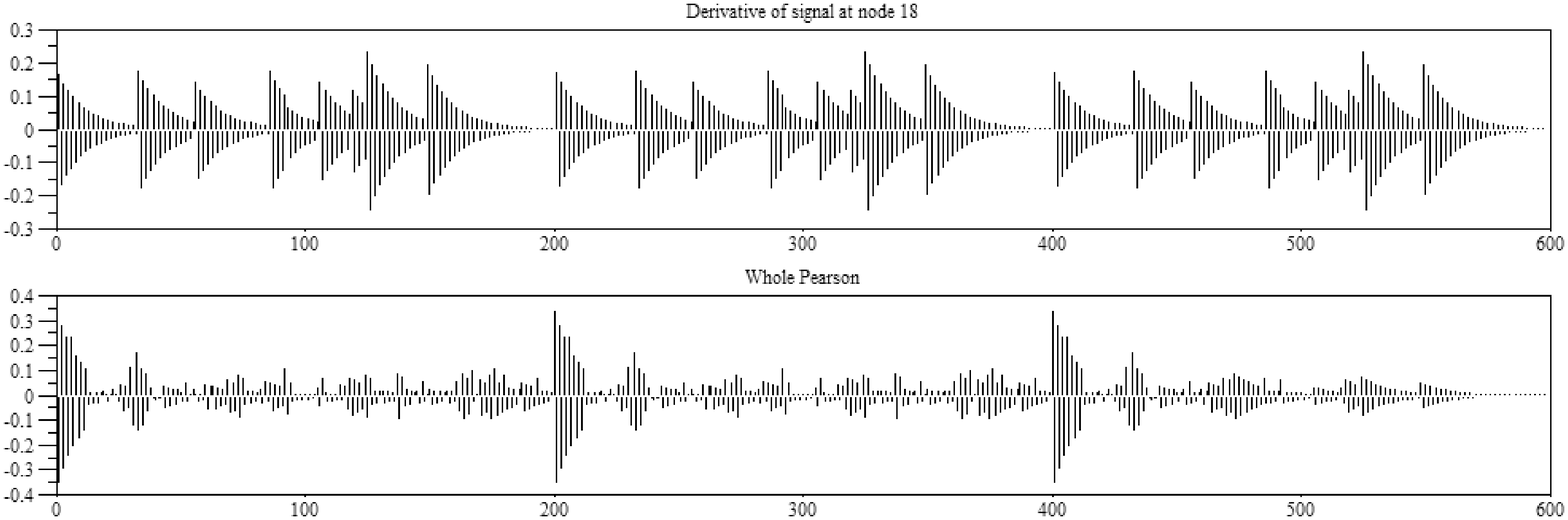}}
  \caption{The original, uniformly random, signal injected into node
           1 of the network in Figure~\ref{fig:net2} and the derivatives
           of the signals at nodes 4, 8, 15 and 21, followed by the 
           respective similarities quantified in terms of the Pearson 
           correlation coefficient between two signals.}
  \label{fig:ex1}
\end{figure*}

Though simple, the previous example allows a number of preliminary
insights about how the connectivity structure of a network can affect
the diffusion of the signal emanating from the source.  The most
important effects of the network structure on the signal arising at a
given node $i$ include: (i) the tendency of the peak lag to reflect
the distance from node $i$ to the source; (ii) the tendency of cycles
and hubs to imply in more disperse and longer system impulse response
at node $i$ (implying less sharp peaks); (iii) the potential
importance of the number of paths of different lengths between the
source and $i$.  In the following section we report a more systematic
investigation about how the diffusion time-varying signals is affected
by distinct network structures.  Because both the magnitude and lag
(and potentially the dispersion) of the peaks seem to be strongly
related to the distance to the source node, special attention is given
to correlating the signals similarity with the respective distances.

\section{Results and Discussion}~\label{sec:exp}

All networks in this article have $N=200$ nodes.  Two different
average degree values have been considered, namely $\left< k \right> =
2$ and $\left< k \right> = 4$. The random signals were obtained for
$\alpha = 0.1$ and basic period $L=200$, repeated $T=8$ times,
implying the total signal duration to be equal to $1600$ time steps.

A total of 1000 realizations were performed for each network
configuration, defined by $N$ and $\left< k \right>$.  Different
random signals were used at each realization, and the source was
always placed at node 1. The activation along time at each node was
Pearson correlated with the original signal in order to obtain a
quantification of the similarity between those two signals.  Three
properties of the similarity signatures --- namely their peak, lag and
entropy --- were then Pearson correlated with the respective distances
to the source node.

\begin{figure*}
  \vspace{0.3cm}
  \centerline{\includegraphics[width=1\linewidth]{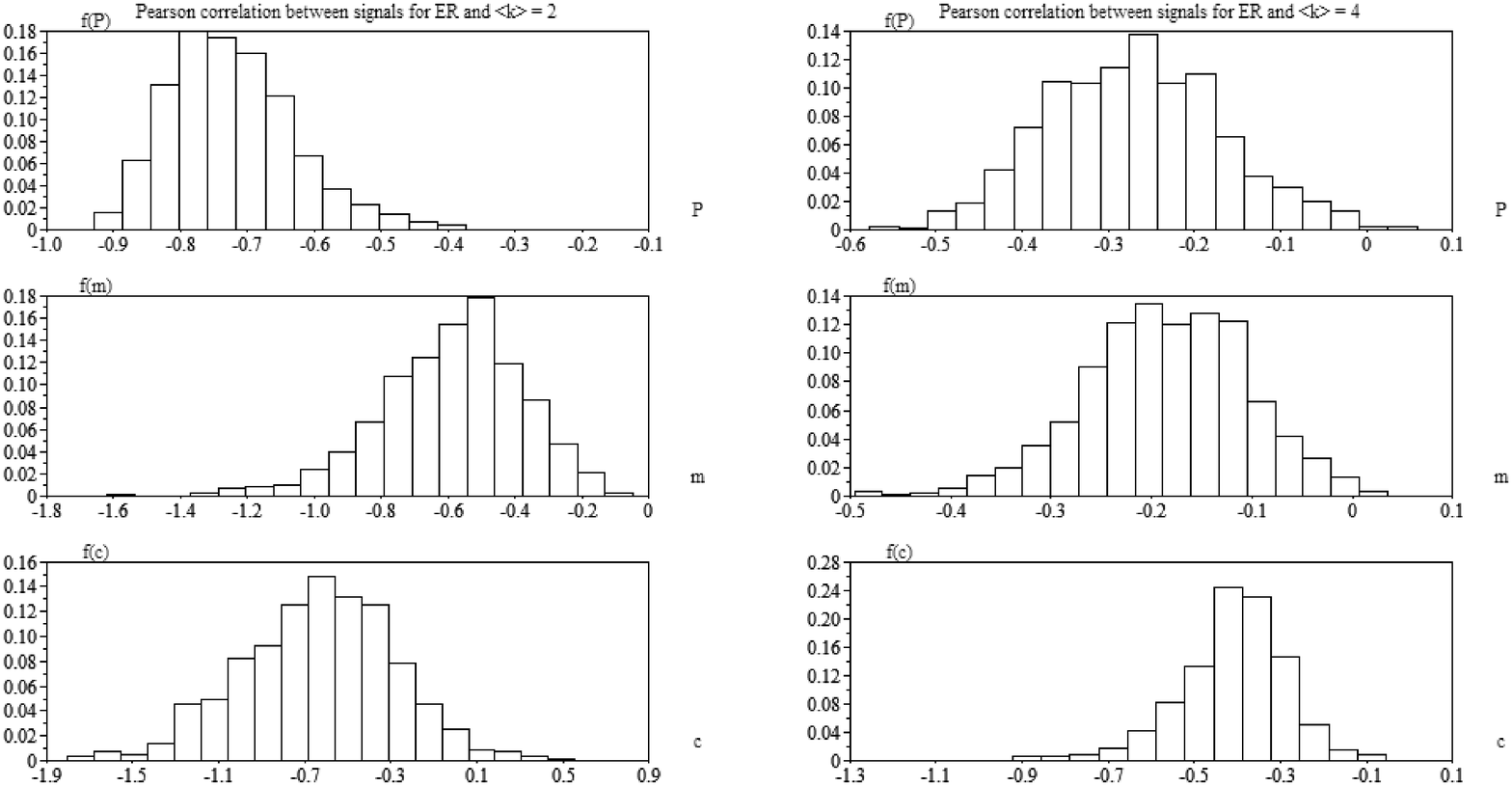}} 
        (a)  \hspace{7cm}  (b) \\
  \centerline{\includegraphics[width=1\linewidth]{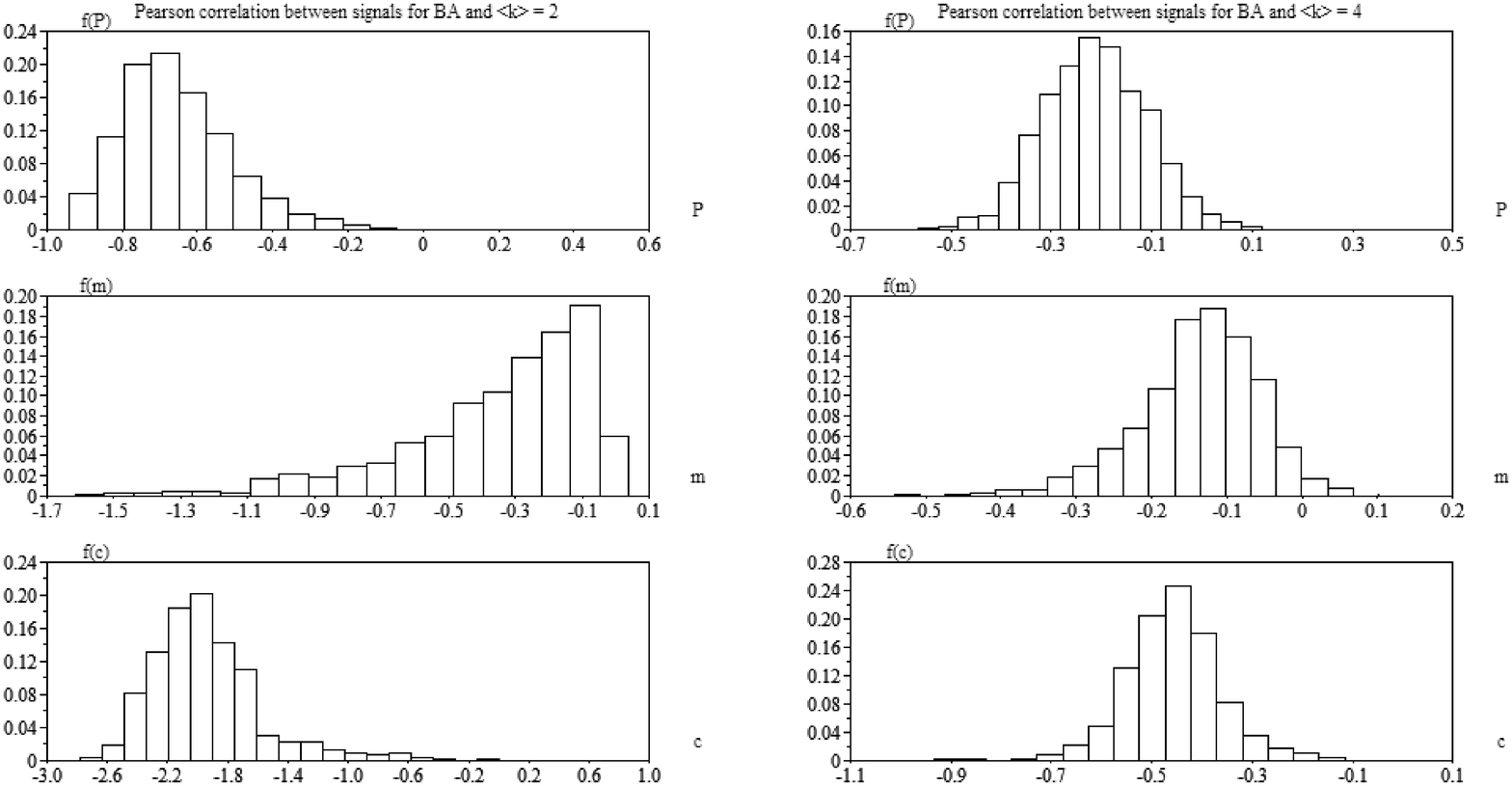}} 
        (c)  \hspace{7cm}  (d) \\
  \caption{The relative frequency histograms describing the
           correlations between the similarity peak magnitude 
           and the distance to the source node for ER and BA
           complex networks with $N = 200$ nodes and 
           $\left<k \right> = 2$ and $4$.  Each vertical group
           shows three relative frequency histograms with respect to the
           Pearson correlation between the peak magnitude and
           the distance to the source (top), as well as the 
           slope (middle) and intercept (bottom) of the respective
           linear regression.}  \label{fig:corrs}
\end{figure*}

Figure~\ref{fig:corrs} refers the correlations between the logarithm
of peak magnitude and the logarithm of the distance to the source
obtained for ER and BA networks with $\left< k \right> = 2$ and
$\left< k \right> = 4$.  Each group of relative frequency histograms
in this figure corresponds to: (i) Pearson correlation coefficient
between the logarithm of the similarity peak magnitude and the
logarithm of the distance to the source node (top histogram, $f(P)$);
(b) slope of the linear regression obtained for the same relationship
(middle histogram, $f(m)$) and (c) respective intercept (bottom
histogram, $f(c)$).

Surprisingly, quite similar relative frequency histograms of the
Pearson correlations have been obtained for ER and BA with the same
average degrees.  This result implies that both networks have similar
properties as far as the presence of the original signal within the
observed derivatives is concerned.  However, the slope and intercept
of the respective linear regressions resulted rather different between
ER and BA with distinct average degrees.  It is also clear from
Figure~\ref{fig:corrs} that the similarity to the activation at the
source node exhibits strong negative correlations with the respective
distance to the source for both ER and BA networks with $\left< k
\right> = 2$.  This is a consequence of the fact that nodes more 
distant to the source tend to be connected to that node through paths
with large dispersion of lengths, hubs and cycles, therefore
contributing to a stronger intermixing of delayed versions of the
original signal.  For $\left< k \right> = 2$, it is clear from the
obtained histograms that the similarity to the original activation at
each node can be predicted with good accuracy from the distance
between that node and the source.  This result establishes a strong
relationship between the structure and the dynamics for signal
diffusion in networks.  The slope of the linear regression between the
similarity and distance to the source is substantially larger for the
ER model than the respective BA counterparts, indicating that the
similarity with the source signal varies more intensely with the
distance to the source in the case of the ER structures.  Very little
variation is observed for BA for both average degrees.  Both the
Pearson correlation and the slope of the linear regression of the
relationship between the similarity and the distance are drastically
reduced with the increase in the average node degree.  This is because
more intense connectivity between nodes tends to create more
alternative paths of different lengths between nodes, hubs and cycles,
contributing to the intermixing the several delayed versions of the
original signal while defining the signals at each node.

\begin{figure*}
  \vspace{0.3cm}
  \centerline{\includegraphics[width=1\linewidth]{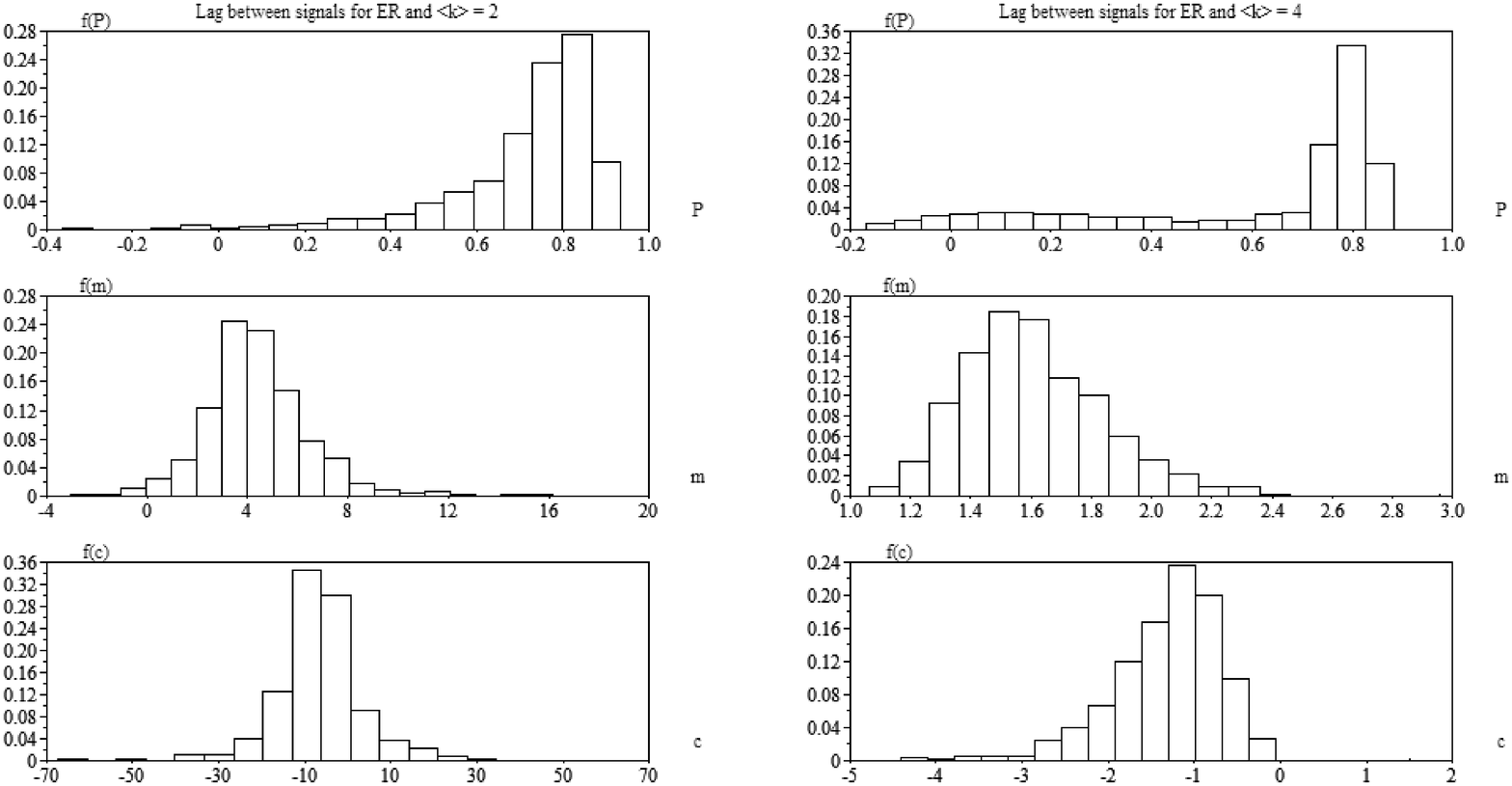}} 
  \centerline{\includegraphics[width=1\linewidth]{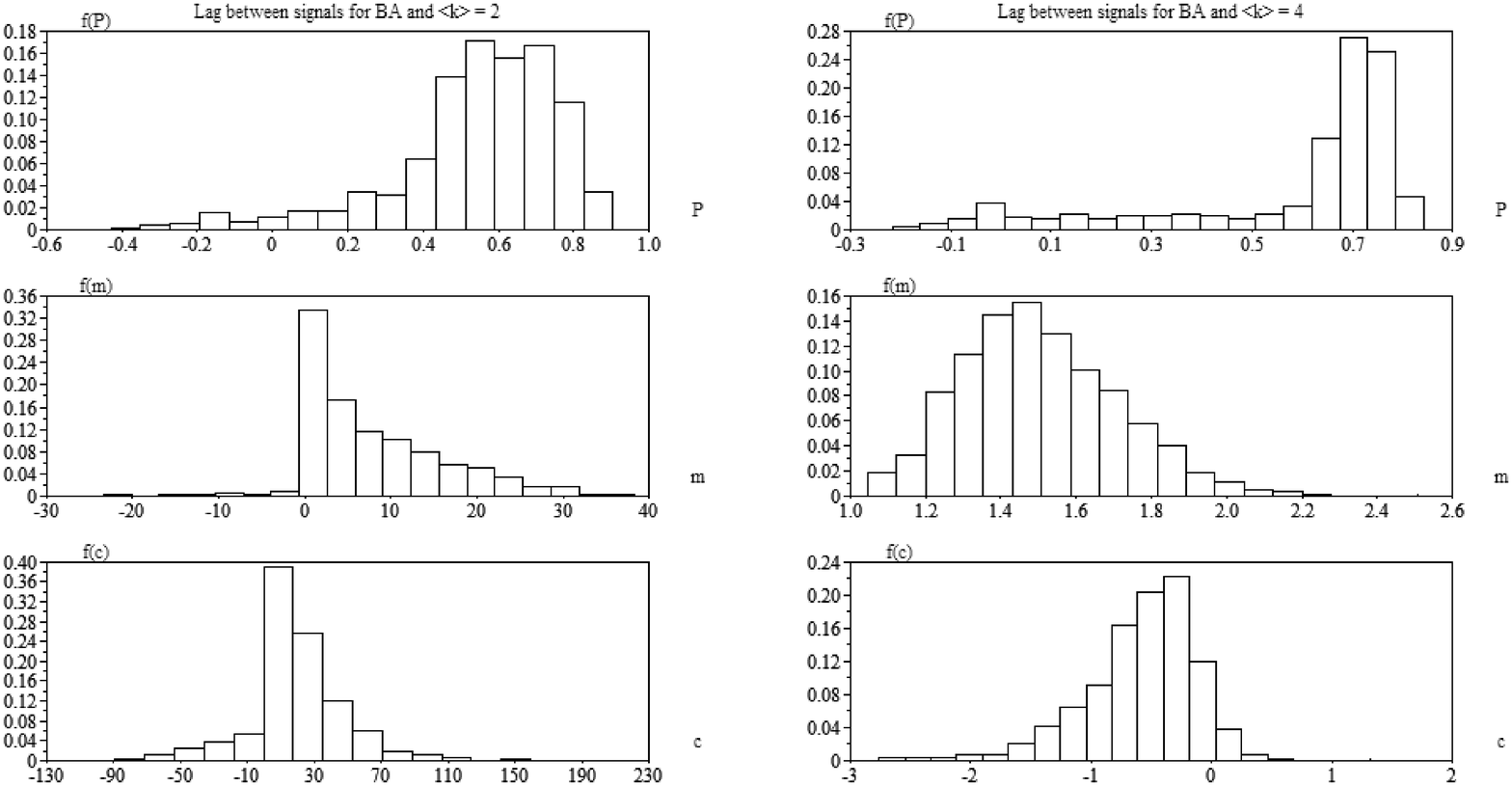}} 
  \caption{The relative frequency histograms describing the
           correlations between the similarity lag 
           and the distance to the source node for ER and BA
           complex networks with $N = 200$ nodes and 
           $\left<k \right> = 2$ and $4$.  Each vertical group
           shows three relative frequency histograms with respect to 
           the Pearson correlation between the peak magnitude and
           the distance to the source (top), as well as the 
           slope (middle) and intercept (bottom) of the respective
           linear regression.}  
  \label{fig:lags}
\end{figure*}

Additional insights about the interplay between the dynamics and
structure in signal diffusion through networks can be gathered by
considering the lag between the detected signal and the original
activation.  Figure~\ref{fig:lags} refers to the correlations between
the lag and the distance between each respective node and the source,
being organized in the same ways as Figure~\ref{fig:corrs}.  Strong
positive correlations can be identified between the lag and the
distance.  However, unlike the results obtained for similarity and
distance (Fig\ref{fig:corrs}), markedly distinct Pearson correlation
coefficient values were obtained for ER and BA for both average
degrees, with smaller correlations being observed in the latter cases.

The relative frequency histograms (not shown in this work) of the
Pearson correlations obtained between the entropies of the similarity
signatures and the respective distances to the source of activation
indicate, in average, that the similarity entropy tends not to
correlate with the distance to the source.  This is a surprising
result because it could be expected that the dispersion would increase
with the distance from the source as a consequence of the additional
intermixings of the signals obtained for longer distances.  The
absence of correlations between the similarity entropy and the
distances to the source is possibly a consequence of the fact that, in
traditional random walks, the signals are allowed to backdiffuse
freely through the whole network.  However, additional investigations
(possibly involving self-avoiding dynamics) are needed in order to
better understand such a lack of correlation.

\section{Concluding Remarks}

Great attention has been focused along the last years on the important
subject of relating structure and dynamics in complex systems.  The
current article reported an investigation about the effect of complex
network structure on the progressive modifications of a time-varying
signal, issued from a specific source node, as it diffuses at the
equilibrium regime.  The undewent modifications were quantified in
terms of three features (peak magnitude, lag and entropy) derived from
the Pearson correlation between the original signal and the derivative
of signals observed at specific nodes.  Though little studied at the
individual level, the linear dynamics of time-varying signal diffusion
provides an interesting case of structure-dynamics in complex network
and is direct and indirectly related to several important types of
natural and artificial linear and non-linear dynamics.  In particular,
relatively little attention has been focused on the problem of
transmission of time-varying signals, which implies the preservation
of the sequence of their constituent parts.  Another distinguishing
feature of the currently reported investigation is that the
relationship between the network structure and dynamics has been
considered with respect to individual pairs of nodes (source and
destination nodes), and not by taking averages through the network.
This is particularly important given that quite distinct dynamics have
been observed in our experiment while considering different pairs of
nodes.  In this work, the consideration of individual pairwise
relationships has been allowed by taking into account not by taking
the average of the observed features through the network, but by
obtainint the relative frequency histograms of the Pearson
correlations calculated between those features and the distance to the
source while considering pairs of nodes involving the source node and
all the other nodes in the networks.

A number of interesting results were obtained with respect to a
systematic investigation (1000 realizations) taking into account ER
and BA structures.  The first important result is that the similarity
peak is almost always obtained near the beginning of each period,
though with varying lags, which indicates that the original signal is
to a certain degree contributing to the variation of the signals
arising at each node.  In addition, a strong negative correlation was
observed between the logarithm of the similarity peak magnitude and
the logarithm of the respective distances to the source node, with
similar values resulting for ER and BA with equivalent average
degrees.  This result indicates that the effect of the original signal
on the variations of the signals observed along time at each node
tends to decrease strong and steadily with the distance to the source.
High positive correlation values were also identified between the peak
lag and the distance to the source node.  However, different
correlation intensities were observed for ER and BA structures in this
case, with larger correlations being obtained for the former type of
network.  So, the lag is more strongly defined by the distance to the
source node in the case of ER networks.  In both cases
(i.e. correlations between peak magnitude and distance to the source
and between peak lag and distance to the source), the correlations
tended to substantially decrease for networks with larger average
degree.  The correlation between the similarity entropy and the
distance to the source tended to present null average, suggesting no
clear relationship between these two features.  Additional
investigations are required in order to better understand the latter
effect.

The above results have immediate implications to many areas, such as
the investigation of the quality of communication channels,
identification of positions in networks, and digital signal
processing.  In the former case, the network connectivity between each
pair of nodes (generalized connectivity or
superedges~\cite{Costa_superedges:2008, Costa_Baggio:2008}) can be
understood as a \emph{communicating channel} between those two nodes.
The methodology reported in this article, as well as related
extensions, can be immediately applied in order to quantify the
fidelity and delay of each of such channels.  

The possibility to identify the position of nodes in non-geographical
networks is directly related to the correlations identified in the
present work between the similarity peak magnitude and lag with the
distance to the source node.  Thus, given a network node and its
respective time-response to a given input signal, it is possible to
estimate its distance to the source node by considering the slope and
intercept of the linear regression between the logarithm of the
magnitude or the lag with respect to the distance to the source.  In
such a case, the source would be acting as a beacon, establishing a
one-coordinate axes through the network (the distances from each node
to the beacon).  It would be particularly interesting to investigate
how the incorporation of additional beacons could contribute to a
one-to-one identification of all the nodes in terms of their distances
to each of several beacons.

The diffusion of time-varying signals in complex networks is a subject
which is intrinsically related to digital signal processing and linear
digital filters (e.g.~\cite{Proakis:2006}).  Indeed, the latter are
structures composed by delays, adders, and multiplications by constant
values. The linear transformation performed on the input signal has
traditionally been investigated in terms of the z-transform, a method
intrinsically suited to treat time-discrete signals.  Thus, it is
possible to represent and studye the diffusion dynamics in complex
networks by using such concepts, i.e. digital filters and
z-transforms.  At the same time, it would be interesting to apply
complex networks concepts, such as shortest path distances between
pairs of nodes, to the area of digital filters.  The main difference
between the work reported in this article and traditional digital
filters is that the latter area typically involves structures which
are typically much smaller and simpler than complex networks.

The reported work also paves the way to a number of further related
investigations. Immediate developments could take into account other
types of networks, correlations with measurements other than the
distance to the source, as well as investigate finite-size effects by
taking into account networks with different numbers of nodes.  In
particular, given that the diffusion effects on the propagated signals
seem to be related to the number of paths of different lengths between
the source and destination nodes, it would be interesting to study the
relationship between such statistics
(e.g.~\cite{Costa_superedges:2008}) and the similarity between the
signals at the source and destinations.  It would also be especially
interesting to repeat the presently reported study to non-linear
self-avoiding dynamics, so as to find out the specific effects of
backward signal propagation allowed by the linear random walks adopted
in the current work.  Even more sophisticated dynamics such as
integrate-and-fire can be investigated from the perspective of the
conceptual and methodological framework proposed in the current work.
As a more leisurely project, it would be possible to transform the
derivative signals into some audio format and hear the sounds arising
at different nodes while the network is stimulated by a given signal
(such as a piece of music) injected at a specific node.

\begin{acknowledgments}
Luciano da F. Costa acknowledges sponsorship from CNPq (301303/2006-1)
and FAPESP (05/00587-5).
\end{acknowledgments}

\bibliography{cn_filters}

\begin{thebibliography}{6}
\expandafter\ifx\csname natexlab\endcsname\relax\def\natexlab#1{#1}\fi
\expandafter\ifx\csname bibnamefont\endcsname\relax
  \def\bibnamefont#1{#1}\fi
\expandafter\ifx\csname bibfnamefont\endcsname\relax
  \def\bibfnamefont#1{#1}\fi
\expandafter\ifx\csname citenamefont\endcsname\relax
  \def\citenamefont#1{#1}\fi
\expandafter\ifx\csname url\endcsname\relax
  \def\url#1{\texttt{#1}}\fi
\expandafter\ifx\csname urlprefix\endcsname\relax\def\urlprefix{URL }\fi
\providecommand{\bibinfo}[2]{#2}
\providecommand{\eprint}[2][]{\url{#2}}

\bibitem[{\citenamefont{Albert and Barab\'asi}(2002)}]{Albert_Barab:2002}
\bibinfo{author}{\bibfnamefont{R.}~\bibnamefont{Albert}} \bibnamefont{and}
  \bibinfo{author}{\bibfnamefont{A.~L.} \bibnamefont{Barab\'asi}},
  \bibinfo{journal}{Rev. Mod. Phys.} \textbf{\bibinfo{volume}{74}},
  \bibinfo{pages}{47} (\bibinfo{year}{2002}).

\bibitem[{\citenamefont{da~F.~Costa et~al.}(2007)\citenamefont{da~F.~Costa,
  Rodrigues, Travieso, and Boas}}]{Costa_surv:2007}
\bibinfo{author}{\bibfnamefont{L.}~\bibnamefont{da~F.~Costa}},
  \bibinfo{author}{\bibfnamefont{F.~A.} \bibnamefont{Rodrigues}},
  \bibinfo{author}{\bibfnamefont{G.}~\bibnamefont{Travieso}}, \bibnamefont{and}
  \bibinfo{author}{\bibfnamefont{P.~R.~V.} \bibnamefont{Boas}},
  \bibinfo{journal}{Advances in Physics} \textbf{\bibinfo{volume}{56}},
  \bibinfo{pages}{167} (\bibinfo{year}{2007}).

\bibitem[{\citenamefont{Newman}(2003)}]{Newman:2003}
\bibinfo{author}{\bibfnamefont{M.~E.~J.} \bibnamefont{Newman}},
  \bibinfo{journal}{SIAM Rev.} \textbf{\bibinfo{volume}{45}},
  \bibinfo{pages}{167} (\bibinfo{year}{2003}).

\bibitem[{\citenamefont{da~Fontoura~Costa and
  Baggio}(2008)}]{Costa_Baggio:2008}
\bibinfo{author}{\bibfnamefont{L.}~\bibnamefont{da~Fontoura~Costa}}
  \bibnamefont{and} \bibinfo{author}{\bibfnamefont{R.}~\bibnamefont{Baggio}}
  (\bibinfo{year}{2008}), \bibinfo{note}{arXiv:0803.2510}.

\bibitem[{\citenamefont{da~Fontoura~Costa}(2008)}]{Costa_superedges:2008}
\bibinfo{author}{\bibfnamefont{L.}~\bibnamefont{da~Fontoura~Costa}}
  (\bibinfo{year}{2008}), \bibinfo{note}{arXiv:0801.4068}.

\bibitem[{\citenamefont{Proakis and Manorakis}(2006)}]{Proakis:2006}
\bibinfo{author}{\bibfnamefont{J.~G.} \bibnamefont{Proakis}} \bibnamefont{and}
  \bibinfo{author}{\bibfnamefont{D.~K.} \bibnamefont{Manorakis}},
  \emph{\bibinfo{title}{Digital Signal Processing}}
  (\bibinfo{publisher}{Prentice Hall}, \bibinfo{year}{2006}).

\end{thebibliography}
\end{document}